\documentclass[12pt]{iopart}
\pdfoutput=1

\usepackage{graphicx}
\usepackage{bm}
\usepackage{epstopdf}
\usepackage{amssymb}
\usepackage{color}
\usepackage[colorlinks=true, letterpaper=true, pdfstartview=FitV, linkcolor=blue, citecolor=blue, urlcolor=blue]{hyperref}
\usepackage{lineno}
\begin{document}

\title[]{Two-dimensional XY ferromagnetism above room temperature in Janus monolayer V$_{2}$XN (X = P, As)}

\author{Wenhui Wan,\textit{$^{a}$} Botao Fu,\textit{$^{b}$} Chang Liu,\textit{$^{c}$} Yanfeng Ge,\textit{$^{a}$} and Yong Liu$^{\ast}$\textit{$^{a}$}}
\address{$^{a}$State Key Laboratory of Metastable Materials Science and Technology $\&$ Key Laboratory for Microstructural Material Physics of Hebei Province, School of Science, Yanshan University, Qinhuangdao, 066004, P. R. China}
\address{$^{b}$College of Physics and Electronic Engineering, Center for Computational Sciences, Sichuan Normal University, Chengdu, China}
\address{$^{c}$Institute for Computational Materials Science, Joint Center for Theoretical Physics (JCTP), School of Physics and Electronics, Henan University, Kaifeng, 475004, China.}
\ead{yongliu@ysu.edu.cn}
\vspace{10pt}
\begin{indented}
\item[]July 2022
\end{indented}

\begin{abstract}
Two-dimensional (2D) XY magnets with easy magnetization planes support the nontrivial topological spin textures whose dissipationless transport is highly desirable for 2D spintronic devices. 
Here, we predicted that Janus monolayer V$_{2}$XN (X = P, As) with a square lattice are 2D-XY ferromagnets by first-principles calculations. 
Both the magnetocrystalline anisotropy and magnetic shape anisotropy favor an in-plane magnetization, leading to an easy magnetization $xy$-plane in Janus monolayer V$_{2}$XN. Resting on the Monte Carlo simulations, we observed the Berezinskii-Kosterlitz-Thouless (BKT) phase transition in monolayer V$_{2}$XN with transition temperature $T_{\rm BKT}$ being above the room temperature. 
Especially, monolayer V$_{2}$AsN has a magnetic anisotropy energy (MAE) of 292.0 $\mu$eV per V atom and a $T_{\rm BKT}$ of 434 K, which is larger than that of monolayer V$_{2}$PN. Moreover, a tensile strain of 5\% can further improve the $T_{\rm BKT}$ of monolayer V$_{2}$XN to be above 500 K.
Our results indicated that Janus monolayer V$_{2}$XN (X = P, As) were candidate materials to realize high-temperature 2D-XY ferromagnetism for spintronics applications.
\end{abstract}

%
\vspace{2pc}

\section{Introduction}
It had been demonstrated that the long-range magnetic order can exist in
two-dimensional (2D) ferromagnetic (FM) materials with an easy magnetization axis including 2D CrI$_{3}$,~\cite{Huang2017} Fe$_{3}$GeTe$_{2}$,~\cite{Fei2018} Cr$_{2}$Ge$_{2}$Te$_{6}$,~\cite{Gong2017} MnSe$_{2}$,~\cite{OHara2018} and VSe$_{2}$.~\cite{Bonilla2018} 
On the other side, 2D FM materials with an easy magnetization plane, which are called 2D-XY FM materials, exhibit a quasi-long-range order that is described by Berezinskii-Kosterlitz-Thouless (BKT) phase transition.~\cite{Kosterlitz1973}
2D-XY FM materials hold the formation of bound pairs of topological nontrivial meron and antimeron spin textures, which are promising for developing high-speed, low-power, and multi-functional spintronic devices and energy-efficient nanoelectronic devices.~\cite{Kosterlitz1974} The material realization of 2D-XY ferromagnets with a high transition temperature ($T_{\rm BKT}$) is a fasting-developing field and remains a challenge.\cite{Wang2021}

2D-XY ferromagnetism had been investigated in 2D hexagonal systems. In the experiment, Tokmachev et al. observed that the sub-monolayer of Eu atoms self-assembled on the silicon surface,~\cite{Tokmachev2021} graphene,~\cite{Sokolov2020} silicene,~\cite{Tokmachev2018} and germanene~\cite{Tokmachev2019} exhibited an easy magnetization plane. However, the corresponding transition temperature $T_{\rm BKT}$ of Eu-based systems are smaller than 20 K.\cite{Tokmachev2021,Sokolov2020,Tokmachev2018,Tokmachev2019} Pinto et al. demonstrated a BKT phase transition in monolayer CrCl$_{3}$ grown on graphene/6H-SiC(0001) substrate with a $T_{\rm BKT}$ of 12.95 K.~\cite{BedoyaPinto2021} 
On the theoretical side, Lu et al.~\cite{Lu2020} and Augustin et al. \cite{Augustin2021} investigated the spin textures and dynamics properties of meron and antimeron in 2D-XY magnet monolayer CrCl$_{3}$. They found that the easy magnetization plane in CrCl$_{3}$ are formed due to competition between the weak out-of-plane single-ion anisotropy and strong in-plane shape anisotropy, as well as the exchange competition between different neighbors, which leads to a low  $T_{\rm BKT}$ of monolayer CrCl$_{3}$.~\cite{Lu2020,Augustin2021} Meanwhile, these topologically non-trivial spin textures are robust to external static magnetic field.~\cite{Lu2020}
Strungaru et al. reported the formation and control of merons and antimerons in monolayer CrCl$_{3}$ by the laser pulse via the two-temperature model.\cite{Strungaru2022}
Besides CrCl$_{3}$, easy-plane magnetism has also been predicted in 
monolayer CrAs with a $T_{\rm BKT}$ of 855 K,~\cite{Ma2020}		
monolayer VSi$_{2}$N$_4$ with a $T_{\rm BKT}$ of 307 K,~\cite{Cui2021} and monolayer FeX$_{2}$ (X = Cl, Br, and I) with a maximum $T_{\rm BKT}$ of 210 K.\cite{Ashton2017} However, the $T_{\rm BKT}$ of these monolayers ~\cite{Ma2020,Cui2021,Ashton2017} has large uncertainty because they were simply estimated by a classical XY model which only considers the nearest-neighboring magnetic coupling and constraints the spin vectors in the $xy$-plane. \cite{PhysRevB.34.292} The 2D-XY ferromagnetism in 2D hexagonal systems motivates the search for easy magnetisation plane in other 2D crystal systems. 
In 2022, Xuan et al. analyzed the 166 stable tetragonal monolayer MX (M = transition metal, X = nonmetal) and found that sufficiently strong M-d/X-p and M-d/M-d couplings can stabilize monolayer MX with the square lattice without ferroelastic transition.~\cite{Xuan2022a}
For example, monolayer VN adopted a square lattice, while monolayer VP and VAs exhibited a rectangular lattice with ferroelasticity.~\cite{Xuan2022,Cheng2021}
However, it had not been reported that 2D square lattices have 2D-XY ferromagnetism. 

Since the synthesis of Janus MoSSe~\cite{Lu2017} and WSSe,~\cite{Lin2020} the symmetry breaking in the vertical direction of 2D Janus structures can introduce extraordinary physics properties.~\cite{Trivedi2020,Du2021,Zhang2020,Li2021,Li2022,Li2021a}
For example, Janus monolayer Cr(I, X)$_{3}$ (X = Cl, Br) and Janus monolayer MnXTe (X = S, Se) exhibit strong spin-orbit coupling (SOC) and Dzyaloshinskii-Moriya interactions (DMIs) due to the structural symmetry breaking in the vertical direction, which leads to the intrinsic domain wall skyrmions or the magnetic-field-induced bimerons.~\cite{Xu2020,Yuan2020} Compared to monolayer 2H-VSe$_{2}$ which is not stable, Janus monolayer 2H-VSSe, 2H-VSeTe, and 2H-VSTe show a structural stability, an easy magnetization plane, and a $T_{\rm BKT}$ of 106 K, 82 K, and 46 K, respectively.~\cite{Dey2020} 
Though monolayer CrI$_3$ is an FM semiconductor with the perpendicular magnetic anisotropy, Janus monolayer Cr$_2$I$_3$Cl$_3$ exhibits in-plane magnetic anisotropy.~\cite{Li2021}
Thus, constructing Janus structures in 2D square systems may be a possible way to realize the easy-plane magnetism. 

In this work, based on V-based tetragonal monolayers, we predicted that Janus monolayer V$_{2}$XN (X = P, As) exhibited a 2D square lattice and 2D-XY ferromagnetism by first-principles calculations. 
We first identified that the FM state was the magnetic ground state for Janus monolayer V$_{2}$XN. Its structural stability was proved by the formation energy, phonon dispersion, thermodynamic studies, and Born criteria. Then, the calculations of magnetic anisotropy energy (MAE) displayed that an easy magnetization plane existed in monolayer V$_{2}$XN. 
The $T_{\rm BKT}$ was estimated by critical behavior of in-plane susceptibility by the Monte Carlo (MC) simulation based on the XXZ model. At last, we found that tensile strain can further enhance the $T_{\rm BKT}$ but have a small influence on the MAE.

\section{Computational method}
The first-principles calculations were performed by the Vienna ab initio simulation package (VASP)~\cite{Kresse1996} with the projector augmented wave (PAW)~\cite{Kresse1999} pseudopotentials and Perdew, Burke, and Ernzerhof (PBE)~\cite{Perdew1996} exchange-correlation functionals. A vacuum layer of 20 \AA\ was adopted to prevent artificial interaction between the adjacent periodic images in the $z$-direction. The kinetic energy cutoff was set to 550 eV.
A $\mathbf{k}$-point mesh of $12\times12\times1$ with the Monkhorst-Pack scheme ~\cite{Monkhorst1976} was used for the Brillouin zone (BZ) of the primitive cell.
The criteria of total energy and atomic force was 10$^{-6}$ eV and 10$^{-3}$ eV/\AA, respectively. To appropriately describe the on-site Coulomb repulsion of the localized $d$ electrons of V atoms, we applied a simplified PBE+U approach proposed by Dudarev et al. \cite{Dudarev1998} An effective Hubbard
term U$_f$ = U - J = 3 eV was adopted for V-d orbitals. In previous works, U$_f$ = 2 $\sim$ 4 eV \cite{Pickett1998,Wang2009,Aryasetiawan2006,Xuan2022} for V-d orbitals had been adopted. We found that the different values of U$_f$ will not change our main conclusions. 
The band structures were further checked by  
Heyd-Scuseria-Ernzerhof (HSE06) hybrid functional.~\cite{Heyd2003} Through the convergence test, the MAE was calculated by a dense $\mathbf{k}$ mesh of $30\times30\times1$ with the SOC effect being included. Phonon dispersion was calculated by a $5\times5\times1$ supercell by the density functional perturbation theory implemented in phonopy codes.~\cite{Togo2015} A $4\times4\times1$ supercell was used to perform the molecular dynamics (MD) simulations in the canonical (NVT) ensemble, up to 5 ps with a time step of 1.0 fs. The MC simulations were performed by a $180\times180\times1$ supercell using MCsolver code.~\cite{Liu2019} We run $2\times10^5$ MC steps per site to reach the thermal equilibrium, followed by $8\times10^5$ MC steps per site for the averaging of magnetization and energies. We have considered different initial spin configurations concluding spin along the one direction, in the one plane or random direction in 3D space. We got consistent final results that the spins in Janus monolayer V$_2$XN (X = P, As) tend to form merons and antimerons when the temperature is below the transition temperature T$_{\rm BKT}$.

\begin{figure}[tbp!]
	\centering
	\includegraphics[width=0.7\textwidth]{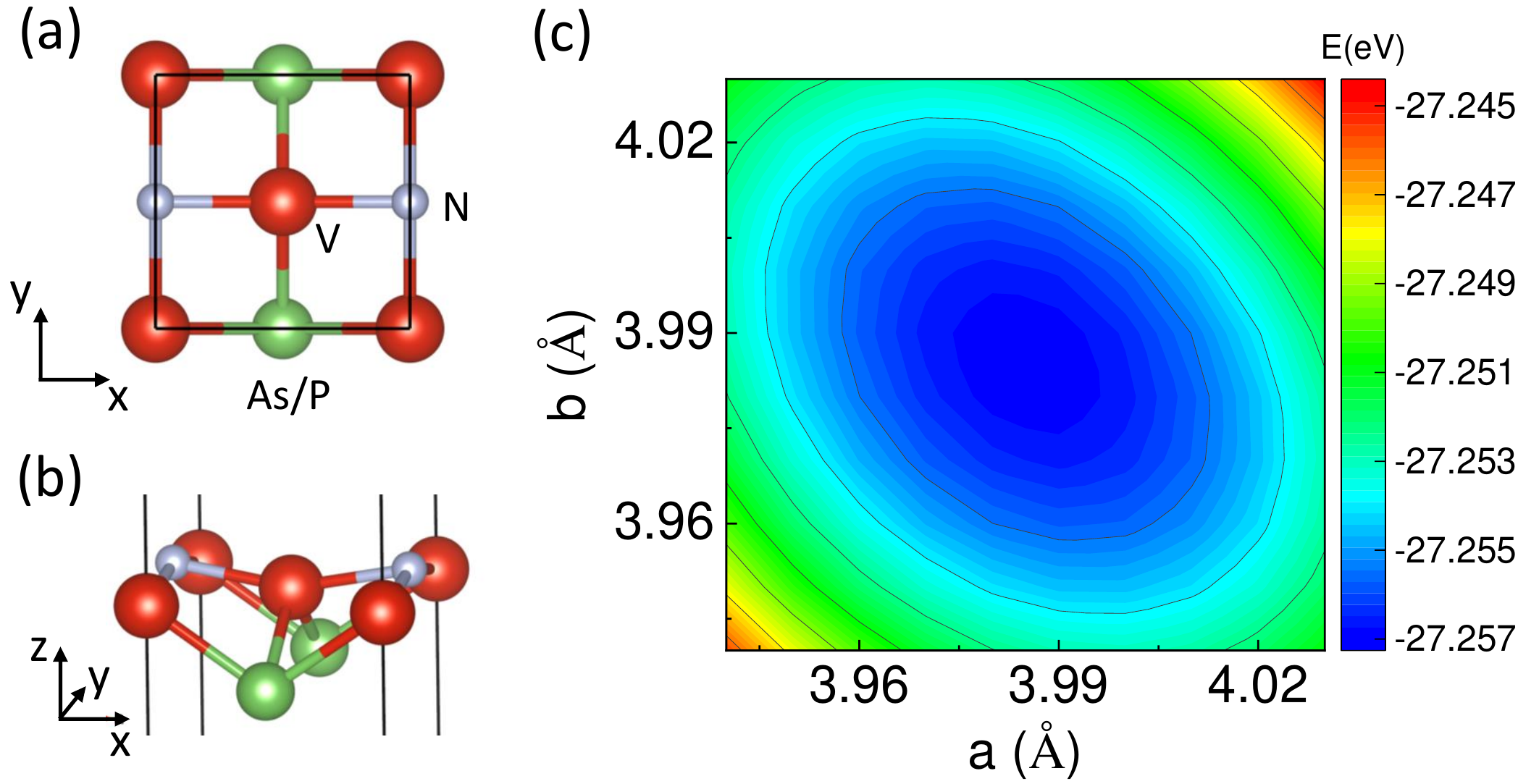}
	\caption{ (a, b) A top view and a side view of the crystal lattice of Janus monolayer V$_{2}$XN (X = P, As). (c) The energy contour of Janus monolayer V$_{2}$PN as a function of lattice constants $a, b$. The lattice with $a=b=3.986$ \AA\ has the lowest energy.}
	\label{wh1}
\end{figure}

\section{Results and discussion}
We first calculated the intrinsic monolayer VN and VP. Monolayer VP adopted a rectangle lattice with the lattice constants of $a = 4.219/4.453$ \AA\ and $b = 3.420/4.443$ \AA\ by PBE/PBE+U calculations. Monolayer VN exhibited a square lattice with the lattice constant of $a = b = 3.653$ \AA\ and $3.990$ \AA\ by PBE and PBE+U calculations, respectively. These results agree well with previous works,~\cite{Xuan2022,Cheng2021,Kuklin2018} indicating the reliability of our calculations. Moreover, monolayer VN adopted a buckling and planar structure in the PBE and PBE+U calculations, respectively [see Fig. S1]. 
Meanwhile, we found that monolayer VN was a non-magnetic (NM) and FM metal in the PBE and PBE+U calculations, respectively. 
The large difference between the results in PBE and PBE+U calculations indicated the necessity to consider the on-site Coulomb repulsion of the localized $d$ electrons in both the structural relaxation and electronic structures of V-based materials. In the following content, we only showed the results of PBE+U calculations.

\begin{figure}[tb!]
	\centering
	\includegraphics[width=0.7\textwidth]{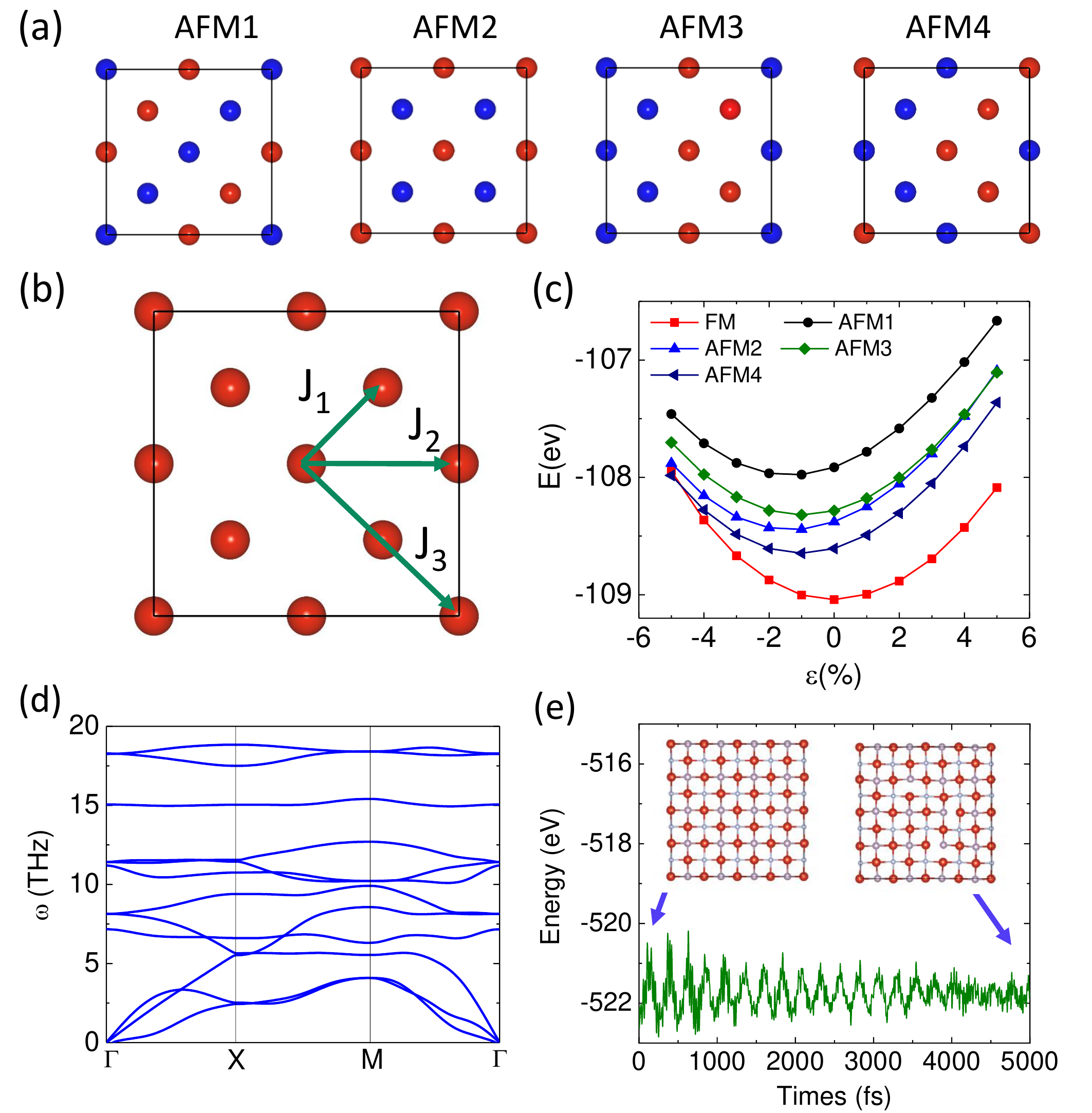}
	\caption{(a) Possible AFM states of monolayer V$_{2}$PN with only V atoms being shown. The red and blue balls represent the V atoms with the up and down spins, respectively.
		(b) FM state and exchange constants J$_{1}$, J$_{2}$, and J$_{3}$.
		(c) The energy of of monolayer V$_{2}$PN in different magnetic states as a function of biaxial strain. (d) The phonon dispersion of monolayer V$_2$PN in the FM state. (e) Evolution of energy of monolayer V$_2$PN as a function of time at T = 300K.}
	\label{wh2}
\end{figure}

Next, we calculated the energy of monolayer V$_{2}$XN (X = P and As) as a function of the lattice constant. Both monolayer V$_{2}$PN and V$_{2}$AsN adopted 2D square lattice as the lowest-energy lattice structure [see Fig.~\ref{wh1}(a-c) and Fig. S2(a)] The corresponding lattice constants are $a = 3.986$ \AA\ and $3.989$ \AA, respectively. The point group and the space group of monolayer V$_{2}$XN are $C_{4v}$ and P4mm (No. 99), respectively. There are two V atoms, an N atom, and an X (X = P and As) atom in the primitive cell. In PBE+U calculations, V atoms are located at the 2c (0.5, 0.0, 0.5) Wyckoff sites. The N and X atoms occupy the 1b (0.5, 0.5, 0.5179) and 1a (0.0, 0.0, 0.4313) Wyckoff sites for monolayer V$_{2}$PN and the 1b (0.000, 0.500, 0.5176) and 1a (0.5, 0.0, 0.4216) Wyckoff sites for monolayer V$_{2}$AsN, respectively. Meanwhile, because both the P and As atoms move away from the V-V plane, the V-V distance in monolayer V$_{2}$XN is smaller than that of intrinsic monolayer VN. As a result, both the N side and X side of monolayer V$_2$XN experience a compressive strain. 
We found that the crystal lattice and magnetic properties of V$_2$PN and V$_2$AsN are similar except that the As atom has a larger buckling height relative to the V-V plane and stronger SOC effect than that of the P atom. We will mainly display the results of monolayer V$_2$PN and give the results of monolayer V$_2$AsN in the end.

To determine the magnetic ground state of monolayer V$_{2}$PN, we considered it in different magnetic states including the NM state, FM state, and four anti-ferromagnetic (AFM) states by a $2\times2\times1$ supercell, as shown in Fig.~\ref{wh2}(a, b). We did not display the NM state as its much higher energy than other magnetic states. We defined the biaxial strain as $\varepsilon = (\frac{a}{a_0}-1)$, where $a$ and $a_0$ are unstrained and strained lattice constants, respectively. 
At $\varepsilon \leq -5 \%$, an FM-to-AFM phase transition occurs [see Fig.~\ref{wh2}(c)]. When $\varepsilon > -5\%$, the FM state is the magnetic ground state for monolayer V$_{2}$PN [see Fig.~\ref{wh2}(c)]. The V-P-V and V-N-V bonding angles between the nearest V atoms of strain-free monolayer V$_{2}$PN are shown in Table~\ref{table1}. The cation-anion-cation angle is close to 90.0$^{\circ}$, which favors an FM coupling between the V atoms according to Goodenough-Kanamori-Anderson superexchange theory.~\cite{Goodenough1955} The energy difference between the FM state and AFM states is 0.39 eV for monolayer V$_{2}$PN, indicating a high transition temperature. We also found that the FM state was
still the ground state of monolayer V$_2$PN for U$_f$ = 2 eV and 4 eV
[see Fig. S3].
According to the Hund’s rules, each V$^{3+}$ ion has two unpaired $d$ electrons with the same spin. The primitive cell with two V ions has a total magnetic moment of 4 $\mu_B$. From the distribution of spin density, we found that the spin density locates mainly around the V atoms. Charge transfers from the V atoms to the N and P atoms due to the electronegativity difference. Based on the Bader charge analysis, the P atom and N atom get 0.90 e and 1.47 e from V atoms, respectively.
The magnetic moments of the V, P, and N ions in a primitive cell are 2.184, -0.224, and -0.255 $\mu_B$, respectively.

To check the stability of monolayer V$_{2}$PN, we calculated the formation energy by $E_{f} = (E_{\rm V_2PN} - 2E_{\rm V} -E_{\rm P} -E_{\rm N} )/4$,
where $E_{\rm V_2PN}$, $E_{\rm V}$, $E_{\rm P}$, and $E_{\rm N}$ are the energies of V$_{2}$PN, V, P, and N atoms in their substance, respectively.
4 is the number of atoms in the unit cell. The $E_{f}$ was estimated as -0.29 eV/atom for monolayer V$_{2}$PN. Meanwhile, monolayer V$_{2}$PN in the FM state was dynamically stable with the absence of imaginary frequency in the phonon dispersion [see Fig.~\ref{wh2}(d)]. Moreover, the MD simulation at 300 K showed that the final crystal lattice of monolayer V$_{2}$PN was maintained without structure frustration, indicating its thermal stability [see Fig.~\ref{wh2}(e)].
As last, the calculated 2D effective elastic constants [see Table~\ref{table1}] satisfy the Born criteria~\cite{Mazdziarz2019} for 2D square lattice:
C$_{11}>0$, C$_{66}>0$, and C$_{11}$> |C$_{12}$|, indicating its mechanical stability. Thus, monolayer V$_2$PN has good structural stability. 

Liu et al. have recently reported the growth of monolayer VN on steel substrate via physical vapor deposition technology. \cite{Liu2021} Cheng et al. proved that monolayer VP could be exfoliated from its bulk counterpart.~\cite{Cheng2021} Similar to the experimental synthesis of Janus monolayer MoSSe~\cite{Lu2017} and WSSe,~\cite{Lin2020} one can replace one side of the P layer of monolayer VP with the N atoms or replace one side of the N layer of monolayer VN with the P atoms to prepare Janus
monolayer V$_2$PN. V-N bonding is shorter and stronger than V–P bonding because of the larger electronegativity between V and N elements. Therefore, it is more favorable to prepare the Janus V$_2$PN from monolayer VP because the replacement of the P is an exothermic reaction.

\begin{table*}[thbh!]
	\small
	\caption{\ Lattice constants $a$(\AA), bonding angle ($^{\circ}$), and 2D effective elastic constants $C_{ij}$ (N/m), the magnetocrystalline anisotropy (MCA), magnetic shape anisotropy (MSA) and MAE of monolayer V$_{2}$XN (X = P, As). The unit of anisotropy energy is $\mu$eV per V atom.}
	\label{table1}
	\begin{tabular*}{\textwidth}{@{\extracolsep{\fill}}ccccccccccccc}
		\hline
		& $a$  & $\angle$V-X-V  & $\angle$V-N-V  & $C_{11}$ &$C_{12}$ & $C_{66}$ & MCA & MSA & MAE \\
		\hline
		V$_{2}$PN    &3.986    & 71.2             & 88.2               & 75.11	 &22.19    &48.36 & 58.8 & 57.8 & 116.6\\
		V$_{2}$AsN   &3.989    & 67.6             & 88.3               & 73.28   &21.53    &44.87 & 230.8 & 61.2 & 292.0  \\
		\hline
	\end{tabular*}
\end{table*}

\begin{figure}[tb!]
	\centering
	\includegraphics[width=0.7\textwidth]{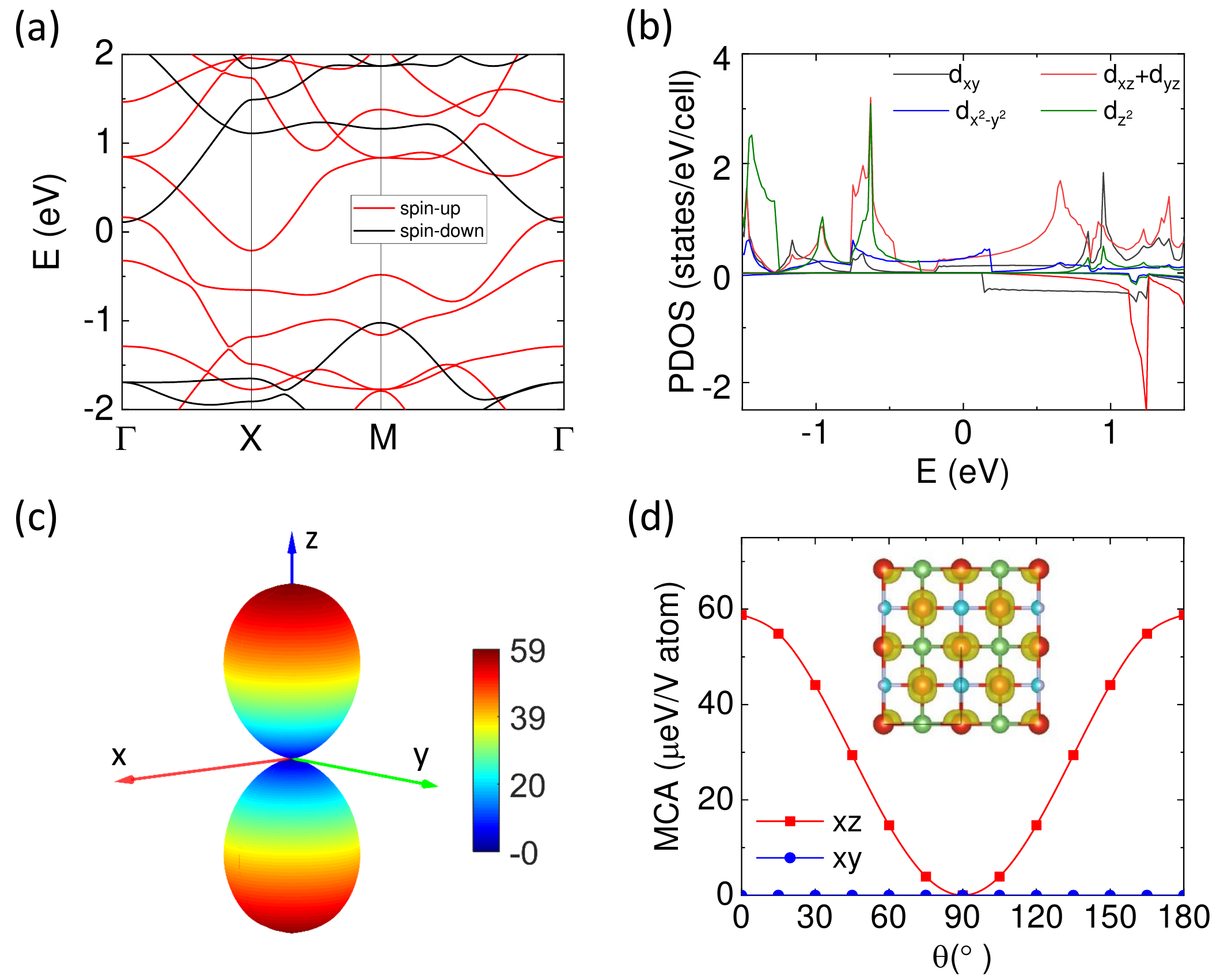}
	\caption{(a) band structure of monolayer V$_2$PN. Red and black lines are spin-up and spin-down bands, respectively.
		(b) Projected density of states (PDOS) of monolayer V$_2$PN. The positive and negative represent the PDOS in the spin-up and spin-down channels, respectively. 
		(c) Angular dependence of the magnetocrystalline anisotropy (MCA) of monolayer V$_2$PN with the direction of magnetization lying on the whole space, 
		(d) The MCA of monolayer V$_{2}$PN along different the orientations of magnetization in the $xz$- and $xy$-plane. The energy corresponding to the magnetization along the $x-$direction is set to be zero. The $\theta$ is the azimuthal angle from the $z$-axis. The inset is the spin density with the isosurface as 0.03 eV/\AA$^3$.}
	\label{wh3}
\end{figure}

The band structure of monolayer V$_{2}$PN is shown in Fig.~\ref{wh3}(a). The spin-up bands cross the Fermi level while spin-down bands have a band gap. 
Thus, monolayer V$_{2}$PN is a half-metal. In the spin-down channel, the valence band maximum (VBM) and the conduction band minimum (CBM) are located at the M point and the $\Gamma$ point of the Brillouin zone (BZ). The energy of the VBM is 1.02 eV lower than the Fermi level, which helps to prevent the thermally excited spin-flip transition. The more accurate HSE06 hybrid functional gave similar band structures [see Fig. S4(a)], indicating the rationality of U$_{f} = 3$ eV in our work. The projected density of states (PDOS) near the Fermi level was mainly contributed by the $d$-orbitals of the V atoms [see Fig. S5]. Due to the $C_{4V}$ symmetry, the $p_x$ and $p_x$ orbitals are degenerate. The $d_{xz}$ and $d_{yz}$ orbitals are degenerate. 
As shown in Fig.~\ref{wh3}(b), the $d_{x^2-y^2}$, $d_{xz}+d_{yz}$, and $d_{xy}$ orbitals of the V atoms dominated the PDOS at the Fermi level in the spin-up channel. 
The $d_{yz}$ and $d_{xy}$ orbitals of the V atoms dominate the PDOS near the VBM and CBM in the spin-down channel, respectively. There is an effective overlap between the PDOS contributed by the $d_{xz}+d_{yz}$ orbitals of V atoms and $p_{x}+p_{y}$ orbitals of P and N atoms around the energy of -3 eV in the spin-up channel and -2 eV in the spin-down channel. The overlap between the PDOS contributed by the V-$d_{x^2-y^2}$ orbital and $p_{z}$ orbitals of P and N atoms are also large in the same energy zone, as shown in Fig. S5. That indicated a strong $p-d$ coupling between the V atom and the P or N atom, which stabilizes the square lattice of monolayer V$_2$PN and excludes the ferroelastic transition.~\cite{Xuan2022a} 

A sizable MAE is significant to maintain magnetic ordering against heat fluctuation. There are two parts of contributions to the MAE. One is the magnetocrystalline anisotropy (MCA) energy ($E_{MCA}$) due to the SOC effect, the other one is the magnetic shape anisotropy (MSA) energy ($E_{MSA}$) due to the magnetic dipole-dipole interaction.
Fig.~\ref{wh3}(c) displayed the MCA of monolayer V$_{2}$PN in the whole space.
The energy of monolayer V$_{2}$PN with magnetization along the in-plane direction is lower than that along the out-of-plane direction. Moreover, the energy difference of monolayer V$_{2}$PN with in-plane magnetization is less than 0.01 $\mu$eV per V atom, thereby being regarded as isotropic in the $xy$-plane, as shown in Fig.~\ref{wh3}(c). In the primitive cell, each V atom has a directional distribution of spin density. However, the spin density of two neighboring V atoms is orthogonal to each other [see the inset of Fig.~\ref{wh3}(d)], which may lead to the in-plane isotropic magnetic properties.
We determined the $E_{MCA}$ of monolayer V$_{2}$PN by the energy difference between the system with magnetization along the $z$-direction and $x$-direction: $E_{MCA} = E^{\rm soc}_{z}-E^{\rm soc}_{x}$.
As shown from Fig.~\ref{wh3}(d), the $E_{MCA}$ is 58.8 $\mu$eV per V atom for monolayer V$_{2}$PN.
The angular dependence of MCA can be expressed by~\cite{Buschow2003}
\begin{equation} \label{fitmae}
	E_{MCA}(\theta)=E_{0}+K_{1} {\rm sin^{2}\theta} + K_{2} \rm sin^{4}\theta.
\end{equation}
Here $\theta$ is the azimuthal angle from the $z$-axis.
$E_{0}$ is a constant. $K_{1}$ and $K_{2}$ are the anisotropy constants.
By fitting the $E_{MCA}$ of $xz$-plane with Eq. \ref{fitmae}, $K_{1}$ and $K_{2}$ were estimated as -58.805 and 0.003 $\mu$eV for monolayer V$_2$PN. The quadratic term in Eq. \ref{fitmae} dominates the MAE of monolayer V$_2$PN.

\begin{figure}[tbh!]
	\centering
	\includegraphics[width=0.7\textwidth]{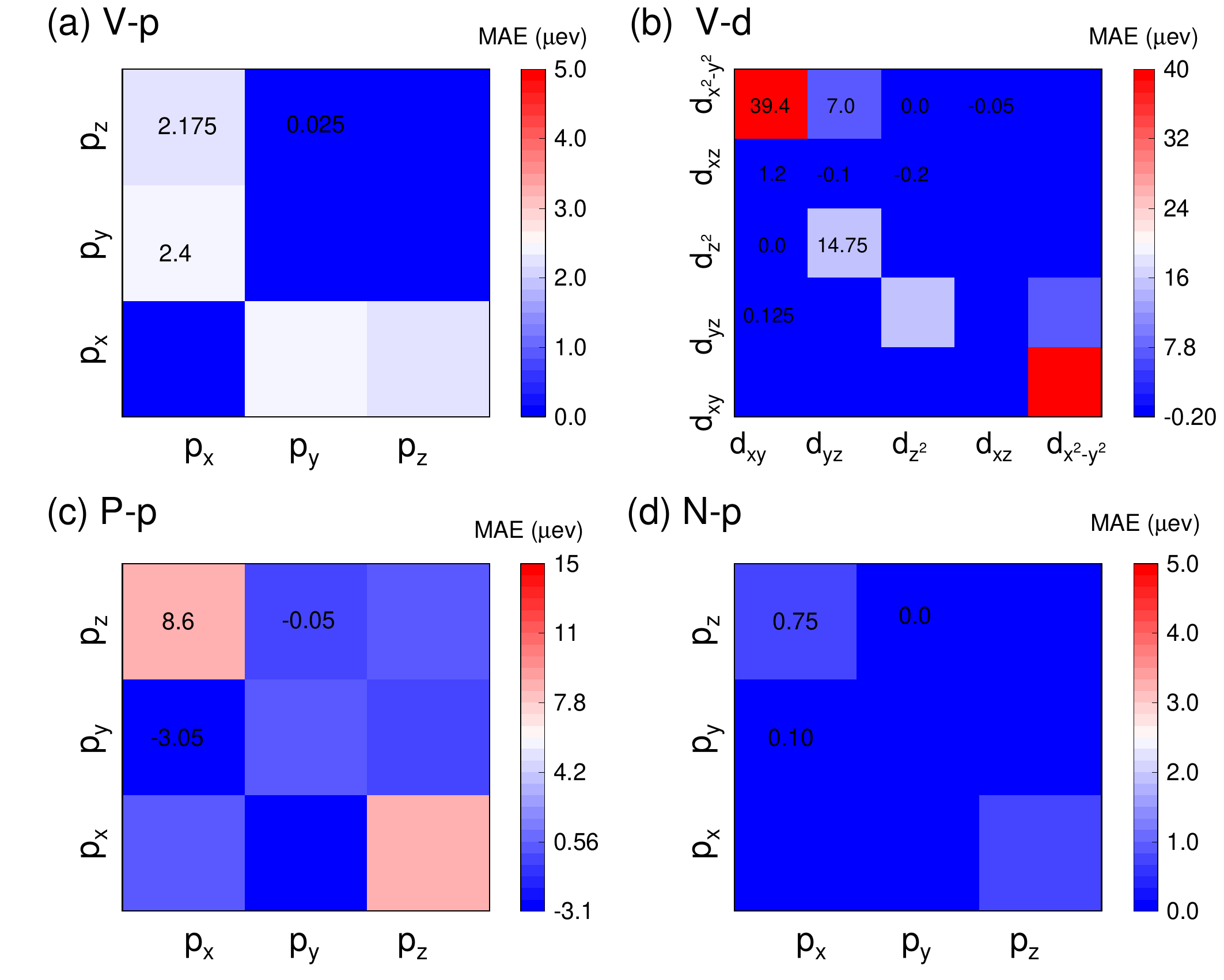}
	\caption{(a) Orbital-resolved MCA of the monolayer V$_2$PN contributed by the hybridization between (a) the $p$ orbitals of the V atoms,  (b) the $d$ orbitals of the V atoms, (c) the $p$ orbitals of the P atoms, and (d) the $p$ orbitals of the N atoms.}
	\label{wh41}
\end{figure}

The orbital-resolved MCA of monolayer V$_{2}$PN was displayed in Fig.~\ref{wh41}. The V-$d$ orbitals make the main contribution to MCA.
Based on the second-order perturbation theory,~\cite{Wang1993} the $E_{MCA} = E^{\rm soc}_{z}-E^{\rm soc}_{x}$ can be expressed: 
\begin{equation} \label{MAE}
	E_{MCA} = \xi^2 (1-2\delta_{\alpha\beta})  \sum_{o,\alpha,u,\beta} \frac{|\left \langle {{o^\alpha}} \right|{\hat L_z}\left| {u^\beta} \right\rangle |^2 -
		|\left\langle {{o^\alpha}} \right|{\hat L_x}\left| {u^\beta} \right\rangle |^2}{\varepsilon^{\alpha}_{u}-\varepsilon^{\beta}_{o}} .  
\end{equation}
where 
$\xi$, $\hat L_{z(x)}$, $\varepsilon_{o}$, and $\varepsilon_{u}$ are the SOC strength, angular momentum operators, the energy levels of occupied states and unoccupied states, respectively. $\alpha$ and $\beta$ are the spin index "+" and "-". $\delta_{\alpha\beta}$ is 1 for $\alpha=\beta$ and 0 for $\alpha\neq\beta$.
The expression of $E_{MCA}$ between our paper and original work~\cite{Wang1993} differs a minus due to the different definitions of $E_{MCA}$.    
Considering the denominator ${\varepsilon_{u}-\varepsilon_{o}}$ of Eq.~\ref{MAE}, the
electronic states near the Fermi level dominate the MCA. Therefore, we did not consider the MCA from the contribution of the occupied state in the spin-down channel due to the large band gap.
Meanwhile, the summation in Eq.~\ref{MAE} is related to the PDOS of occupied and unoccupied states.
We classified the $d$ orbitals by magnetic quantum number $m$.
We used the $d_{m=0}$, $d_{|m|=1}$, and $d_{|m|=2}$ to represent $d_{z^2}$, \{$d_{xz}$, $d_{yz}$\}, and \{$d_{xy},d_{x^2-y^2}$\}, respectively.
The matrix element in the numerator of Eq.~\ref{MAE} is nonzero for
$\left\langle {m} \right|{\hat L_z}\left| {m} \right\rangle $ and
$\left\langle {{m}} \right|{\hat L_x}\left| {m\pm1} \right\rangle $.
Combining the Fig.~\ref{wh3}(b) and ~\ref{wh41}(b), the $E_{MCA}$ is mainly and positively contributed by
$\left\langle {d^{+}_{x^2-y^2}} \right|{\hat L_z}\left| d^{-}_{xy} \right\rangle $ due to that the $d_{xy}$ orbital in the CBM of spin-down bands is close to the Fermi level. 
The hybridization $\left\langle {d^{+}_{x^2-y^2}} \right|{\hat L_x}\left| d^{+}_{yz} \right\rangle $ and $\left\langle {d^{+}_{z^2}} \right|{\hat L_x}\left| d^{+}_{yz} \right\rangle $ also make positive contributions to the $E_{MCA}$.
All these parts lead to a positive $E_{MCA}$ for monolayer V$_{2}$PN, which favors  the in-plane magnetization of the V atom.

Compared to MCA, the magnetic dipole-dipole interaction usually favors the magnetization along the elongated direction of 1D or 2D materials.\cite{Johnston2016}
We estimated the MSA energy $E_{MSA}=E^{dipole}_{z}-E^{dipole}_{x}$ of monolayer V$_2$PN with a classical magnetic dipole-dipole interaction model [see Fig. S6(a)]. The $E_{MSA}$ is also positive, indicating that shape anisotropy also favors magnetization along the in-plane direction. Meanwhile, the $E^{dipole}$ is isotropic in the $xy$-plane [see Fig. S6(b)]. Thus, the MAE =$E_{MCA}+E_{MSA}$ of monolayer V$_{2}$PN is isotropic in the $xy$-plane, indicating an easy magnetization $xy$-plane. Moreover, the mechanism of forming the easy-plane ferromagnetism in monolayer V$_{2}$PN is different from monolayer CrCl$_3$. Lu et al. have reported that the net effect of competition between the strong in-plane MSA and weak out-of-plane MCA leads to the easy magnetization plane of monolayer CrCl$_3$.~\cite{Lu2020}

Compared to Janus monolayer V$_2$PN, the $E_{MCA}$ and $E_{MSA}$ of monolayer VN are -140.1 $\mu$eV per V atom and 36.9 $\mu$eV per V atom, respectively  [see Fig. S1(c)]. Thus, the MAE of monolayer VN is negative, indicating that monolayer VN exhibit an easy magnetization $z$-axis.
Our result is contrasted with previous Kuklin's work,~\cite{Kuklin2018} due to that we predicted a planar square lattice by PBE+U calculations while Kuklin et al. predicted a bucking square lattice by PBE calculations~\cite{Kuklin2018} [see Fig. S1 (a, b)].
The $d_{|m|=1}$ = \{$d_{xz}, d_{yz}$\} orbitals in the spin-up channel dominated the PDOS around the Fermi level [see Fig. S1(c)]. The orbital-resolved MAE shows that the negative contribution of
$\left\langle d^{+}_{|m|=1} \right|{\hat L_z}\left| d^{+}_{|m|=1} \right\rangle $
is the main reason for the negative MCA of monolayer VN. As a result, the spins in monolayer VN prefers to be oriented along the $z$-direction through the normal FM phase transition. 

Since Janus monolayer V$_2$PN is a 2D-XY ferromagnet, it will occur a BKT phase transition at low temperatures,~\cite{Kosterlitz1973} similar to monolayer CrCl$_3$.~\cite{Lu2020} 
Considering that a true spin is a 3D vector, we adopted the XXZ model to estimate the transition temperature $T_{\rm BKT}$:
\begin{equation} \label{Heisenberg}
	H = - \sum_{\left\langle i, j \right\rangle }  [ J_{x} (S^{x}_{i} S^{x}_{j} + S^{y}_{i} S^{y}_{j}) + J_{z} S^{z}_{i} S^{z}_{j} ]  - D \sum_{i} (S^{z}_{i})^2,	
\end{equation}
where $\bf{S}_{i}$ is the spin vector on site $i$ and is normalized to 1.
$D$ is the single-ion anisotropy.
We considered the exchange constants $J$ up to the third nearest neighbors [see Fig.~\ref{wh2}(b)].
The energy of the FM and AFM states with magnetization along the $z$-axis were expressed as
\begin{equation} \label{fmz}
	E^{z}_{\rm FM}   = E_0 - 2J_{1,z} - 2 J_{2,z} - 2 J_{3,z} - D, \\
\end{equation}
\begin{equation}
	E^{z}_{\rm AFM1} = E_0 + 2 J_{2,z} - 2 J_{3,z} - D,  \\
\end{equation}
\begin{equation}
	E^{z}_{\rm AFM2} = E_0 + 2 J_{1,z} -2 J_{2,z} - 2 J_{3,z} - D, \\
\end{equation}
\begin{equation} \label{afm3z}
	E^{z}_{\rm AFM3} = E_0 + 2 J_{3,z} -D.  
\end{equation}

The energy corresponding to the magnetization along the $x$-axis can be obtained by replacing $J_z$ by $J_x$ and dropping 
the single-ion anisotropy $D$ in above equations. 
We got eight equations to solve the exchange constants and single-ion anisotropy.   
The energies of different magnetic states can be obtained by first-principles calculations.
By substituting the energies into Eq.~\ref{fmz} to \ref{afm3z}, we obtained the $J$ and $D$, as shown in Table~\ref{table2}. All the exchange constants are positive indicating the FM coupling between the V ions at different neighbors.
The $J_{3}$ is much smaller than $J_{1}$ and $J_{2}$. The $D$ is negative, which indicates that spins of V ions prefer to be in-plane rather than out-of-plane in monolayer V$_2$PN. 

\begin{table}[tbh!]
	\small
	\caption{\ The exchange constants $J$ (meV) and single-ion anisotropy $D$ (meV) of monolayer V$_{2}$PX (X = P, As).}
	\label{table2}
	\begin{tabular*}{0.8\textwidth}{@{\extracolsep{\fill}}cccccccccccc}
		\hline
		&$J_{1,x}$  &$J_{1,z}$  &$J_{2,x}$  &$J_{2,z}$   &$J_{3,x}$  &$J_{3,z}$ & D \\
		\hline
		V$_{2}$PN   &21.12      &21.09     &25.44      &25.43   &2.833     &2.834    & -0.007    \\
		V$_{2}$AsN  &25.26      &25.16     &23.21      &23.18   &6.24      &6.25     & -0.008    \\
		\hline
	\end{tabular*}
\end{table}

We first made a qualitative estimation of $T_{\rm BKT}$ of monolayer V$_2$PN.
With 2D square lattice with the nearest-neighboring exchange constant $J_1$ being considered,   
the scaling and renormalization-group calculations of XXZ model \cite{Fischer1993}  gave the following form of T$_{BKT}$:
\begin{equation} \label{re}
	T_{\rm BKT}	= \frac{4 \pi J_{1,x}}{A+{\rm In}[(1-\lambda)^{-1}]}.  
\end{equation}
with $\lambda = J_{1,z}/J_{1,x}$. Cuccoli et al. predicted that the $T_{\rm BKT}= 0.546 J_{1,x}/k_B$ for 2D square lattice with $\lambda = 0.99$ by MC calculations.\cite{Cuccoli1995}
So we could estimate the constant $A$ in Eq.\ref{re} to be about 18.410. 
On the other side, Gouva et al. proposed that the T$_{BKT}$ was scaled by $(1 + 2J_2/J_1)$ by including the second nearest-neighboring exchange constant $J_2$.~\cite{Gouvea2002} 
By substituting the exchange constants from Table~\ref{table2} into the Eq. \ref{re}, we predicted the $T_{\rm BKT}$ of monolayer V$_2$PN to be about 419.0 K which has exceeded the room temperature. 

\begin{figure}[tbh!]
	\centering
	\includegraphics[width=0.7\textwidth]{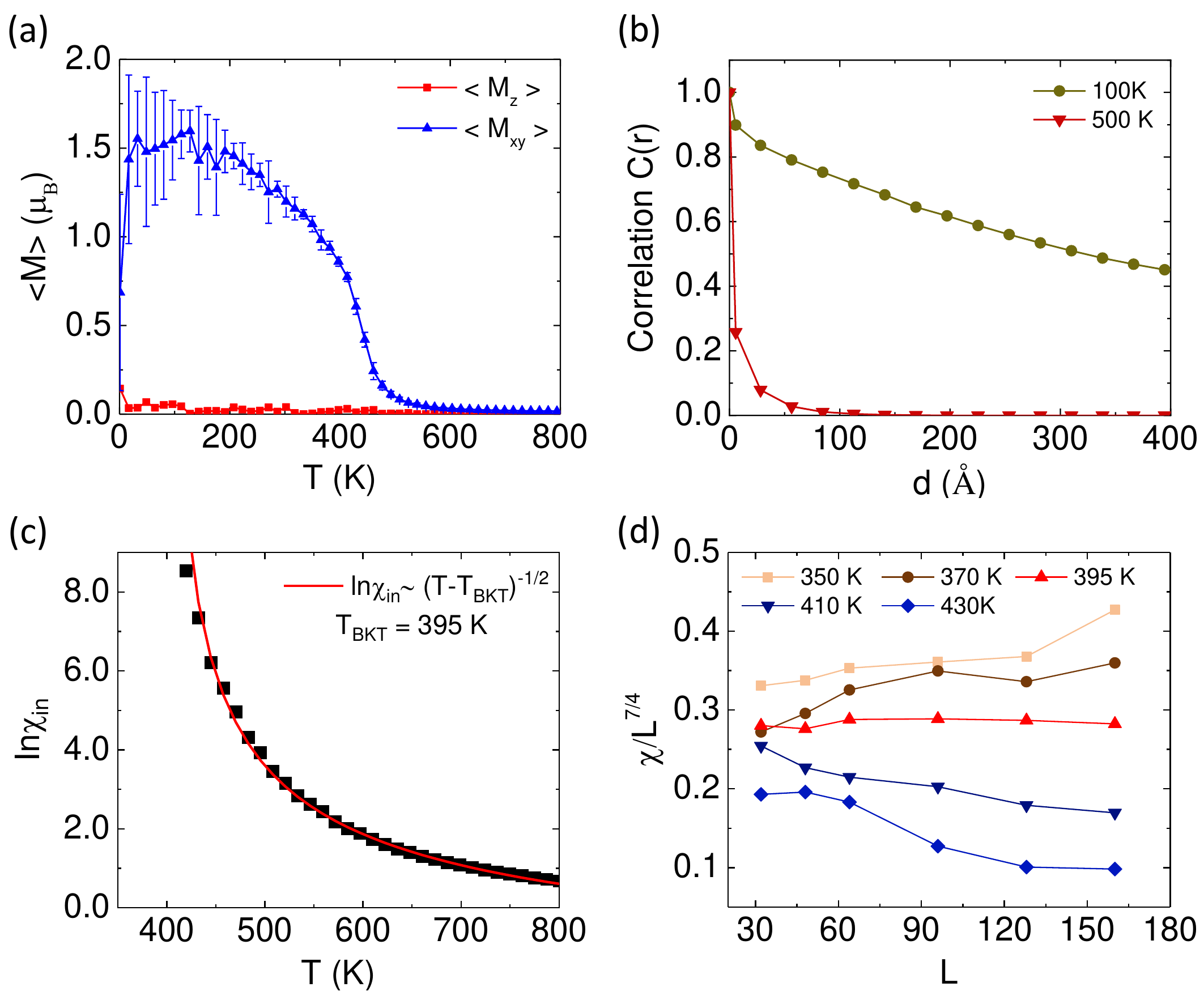}
	\caption{(a) Temperature dependence of averaged magnetization $\left\langle M_{xy} \right\rangle $ and $\left\langle M_z \right\rangle$ in monolayer V$_2$PN. The error bars are the standard deviation. (b) The spin-spin correlation function $C(r)$ of in monolayer V$_2$PN at T = 100 K and 500 K.
		(c) In-plane susceptibility $\chi_{\rm in}$ of monolayer V$_2$PN as a function of temperature.
		(e) $\chi_{\rm in}/L^{7/4}$ vs lattice size $L$ of monolayer V$_2$PN at different temperatures.
	}
	\label{wh4}
\end{figure}

Next, we performed exact MC simulations.
The out-of-plane averaged magnetic moments $\left\langle M_{z} \right\rangle$
is much smaller than in-plane averaged magnetic moments $\left\langle M_{xy} \right\rangle$ at any temperature [see Fig.~\ref{wh4}(a)].
The in-plane magnetization exhibit a large fluctuation at low temperatures, which can be explained by the emergence of meron and antimeron spin textures.~\cite{Lu2020}
The spin-spin correlation~\cite{Lu2020} $C(r)= \left\langle {\bf S}(0) \cdot {\bf S}(r) \right\rangle $ of V$_{2}$PN
occurs an exponential decay to zero at a high temperature (500 K) at which V$_{2}$PN is in the paramagnetic phase [see Fig.~\ref{wh4}(b)].
At a low temperature (100 K), $C(r)$ exhibits an algebraic decay continuously and indicates a quasi-long-range order through the BKT phase transition.~\cite{Kosterlitz1973}

The $T_{\rm BKT}$ can not be determined by specific heat $C_V$ data as it displays not a divergence but a maximum value at a temperature above $T_{\rm BKT}$ for a 2D-XY ferromagnet.~\cite{Gupta1992,Cuccoli1995}
Cuccoli. et al~\cite{Cuccoli1995} defined the susceptibility as the average of the squared magnetization,
\begin{equation} \label{susceptibility3}
	\chi^{\alpha\alpha}= \frac{1}{N} <    ( \sum\nolimits_{i = 1}^N S^\alpha_i )^2 >,
\end{equation}
and in-plane susceptibility as $\chi_{\rm in} = (\chi^{xx}+\chi^{yy})/2$. The $N$ is the number of magnetic moments. $\left\langle ... \right\rangle$ denotes the time average of corresponding components.
Such a definition of $\chi_{in}$ retains its value in investigating the divergence of $\chi_{in}$ also for finite lattice simulations.~\cite{Cuccoli1995}
It has been proved that $\chi_{in}$ diverges exponentially as $T_{c}$ is approached from above ($T \rightarrow T^+_{\rm BKT}$):
\begin{equation} \label{susceptibility1}
	\chi_{\rm in} \sim e^{b (T-T_{\rm BKT})^{-1/2}},
\end{equation}
where $b$ is a non-universal constant. $\chi_{in}$ satisfies the form of
\begin{equation} \label{susceptibility2}
	\chi_{\rm in} \sim L^{2-\eta}
\end{equation}
for $T \leq$ $T_{\rm BKT}$,
where $L$ is the size of the supercell. The exponent $\eta$ is a function of temperature $T$ and satisfies $\eta = 1/4$ at $T_{BKT}$.~\cite{Cuccoli1995}

We fitted the $\chi_{\rm in}$ with Eq.~\ref{susceptibility1} and estimated the $T_{\rm BKT}$ of monolayer V$_2$PN to be about 395 K
[see Fig.~\ref{wh4}(c)], which is much higher than $T_{\rm BKT} = 12.95$ K of synthesized monolayer CrCl$_{3}$.~\cite{BedoyaPinto2021,Lu2020}
To justify this prediction of $T_{\rm BKT}$, we performed the MC simulations by supercells with different sizes $L$.
We explored the finite-size scaling behavior of in-plane susceptibility $\chi_{in}$ 
by calculating the $\chi_{in}/L^{7/4}$ as a function of supercell size $L$, as shown in Fig.~\ref{wh4}(d).
At $T_{\rm BKT} = 395$ K, the data $\chi_{in}/L^{7/4}$ is insensitive to size $L$, indicating the behavior of $\chi_{in} \sim L^{7/4}$.
That is consistent with the exponent $\eta=1/4$ at $T_{\rm BKT}$ [see Eq. \ref{susceptibility2}].~\cite{Cuccoli1995} 
Meanwhile, the $T_{\rm BKT}$ is also consistent with the aforementioned rough estimation of $T_{\rm BKT}$ by the renormalization-group calculations [see Eq.~\ref{re}].~\cite{Wang2021} Thus, the estimation of $T_{\rm BKT}$ is reliable.

\begin{figure}[tbh!]
	\centering
	\includegraphics[width=0.7\textwidth]{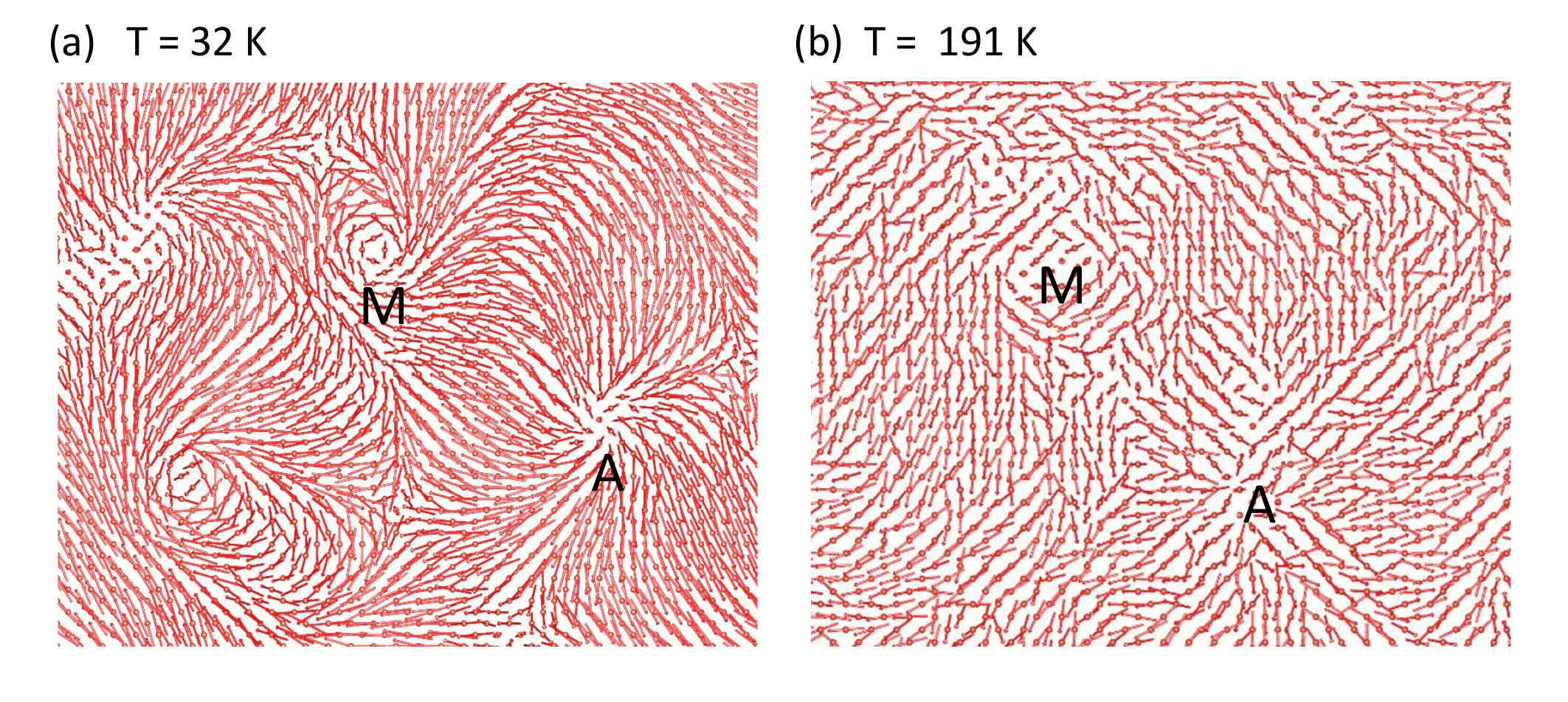}
	\caption{The top view of the real-space arrangement of magnetic moments of V$_2$PN after MC simulation at (a) T = 32 K and (b) T = 191 K. The length of the arrows represents the in-plane component of magnetic moments. The position of meron (M) and antimeron (A) are labeled.}
	\label{wh5}
\end{figure}

The representative real-space arrangement of magnetic moments after MC simulation is shown in Fig. \ref{wh5} and Fig. S7 (a).
The meron and antimeron spin textures were observed for temperature $T < T_{\rm BKT}$ [see Fig. \ref{wh5}(a)].
It was expected that the transport of these topological nontrivial spin textures is dispersionless at low temperatures,
which can be applied in energy-efficient electronic devices.~\cite{Lu2020}
The magnetic moments prefer to lie in-plane around spin defects but along the out-of-plane direction in the core of spin textures [see Fig. S7 (c ,d)], 
which is consistent with Wysin's prediction that perfect in-plane meron and antimeron is unstable and will develop out-of-plane spin component for 2D square lattices with $\lambda = J_z/J_x > 0.704$.~\cite{Wysin1994} Actually, the $\lambda$ has reached 0.99 for monolayer V$_2$PN [see Table \ref{table2}].
As temperature increases, the spin structures of spin textures were disturbed by thermal energy. The number of merons or antimerons was decreased by thermal fluctuations [see Fig. \ref{wh5}(b) and Fig. S7(b)].
At $T \geq $ $T_{\rm BKT}$, we only observed the paramagnetic states.

We expanded the current calculations to monolayer V$_{2}$AsN. The FM state is the magnetic ground state for monolayer V$_{2}$AsN with U$_f$ of $2 \sim 4$ eV [see Fig. S2 (b) and S3 (b)]. 
The magnetic moments of the V, As, and N ions in a primitive cell are 2.236, -0.267, and -0.259 $\mu_B$, respectively.
Its formation energy $E_{f}$ is -1.20 eV/atom. Its structural stability can be seen from the Fig. S2 (c, d) and Table ~\ref{table1}. The electronic structures indicate that monolayer V$_{2}$AsN is also a half metal [see Fig. S2 (e, f) and S4 (b)]. The CBM of monolayer V$_{2}$AsN in the spin-down channel is more far away from the Fermi level than that of monolayer V$_{2}$PN. 
The estimation of MCA, MSA, MAE, exchange constants and $T_{\rm BKT}$ are shown in Fig. S6(a), Fig. S8 (a, b), Table~\ref{table1} and Table~\ref{table2}. 
Due to the larger SOC effect, monolayer V$_{2}$AsN possesses a larger MAE of 292 $\mu$eV per V atom, stronger exchange constants $J$, and a higher $T_{\rm BKT}$ of 434 K than that of monolayer V$_2$PN. 

\begin{figure}[htb!]
	\centering
	\includegraphics[width=0.7\textwidth]{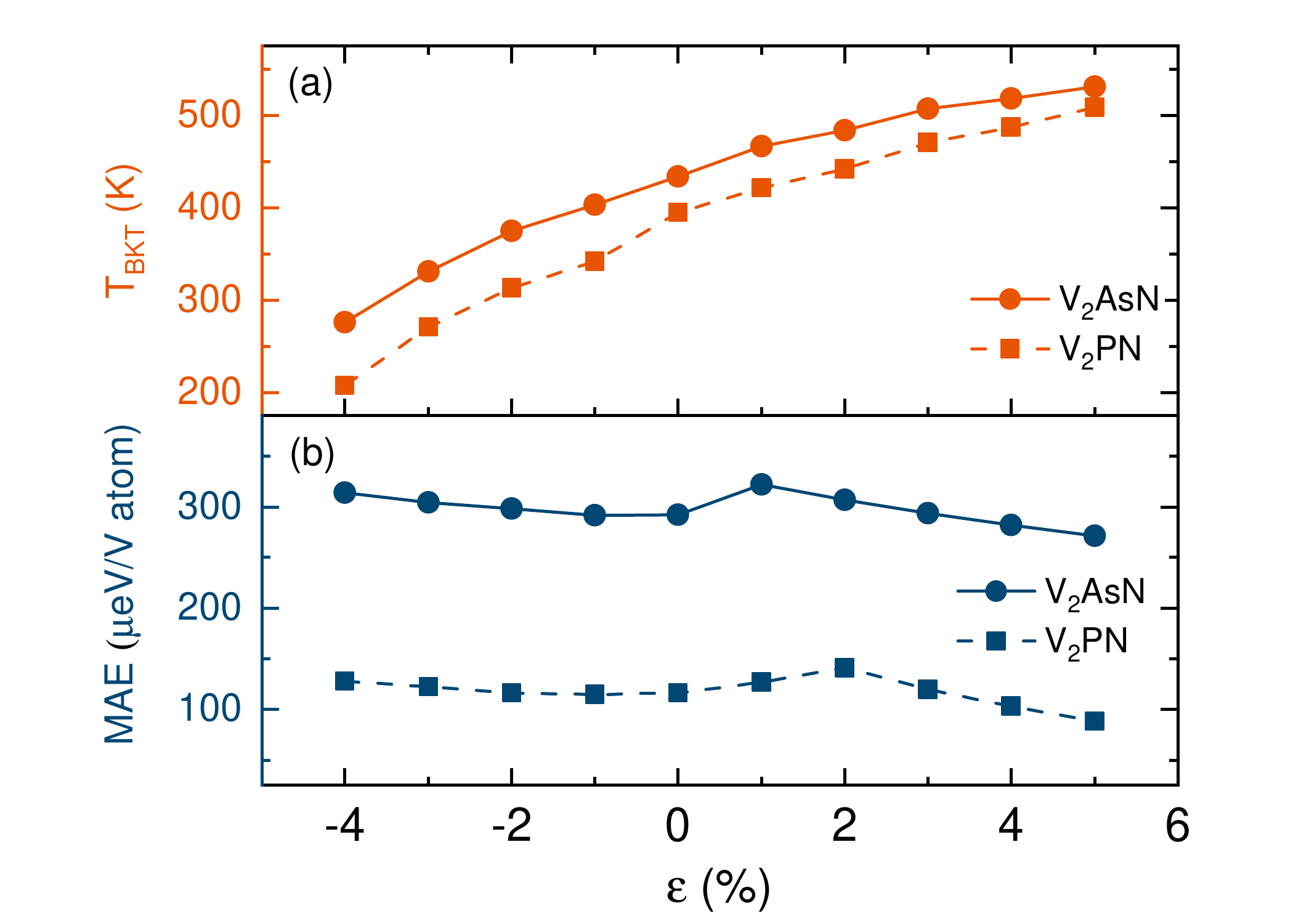}
	\caption{The (a) transition temperature $T_{\rm BKT}$ and (b) MAEs of monolayer V$_2$PN and V$_2$AsN as function of strain.}
	\label{wh6}
\end{figure}

At last, we investigated the strain effect on the magnetic properties of monolayer V$_2$XN (X = P, As).
As the FM state has higher energy than AFM states at $\varepsilon = -5\%$ [see Fig. \ref{wh2}(c)], we only varied the biaxial strain $\varepsilon$ from -4\% to 5\%.
We found that tensile strain increases the bond angles $\angle$V-X-V and$\angle$V-N-V in monolayer V$_2$XN (X = P, As), making them closer to 90$^{\circ}$. The FM coupling and the exchange constant $J$ between V atoms was enhanced by tensile strain. As a result, the $T_{\rm BKT}$ of monolayer V$_2$PN and V$_2$AsN can be increased to 508.7 K and 531.3 K by a tensile strain of 5\%, respectively. 
That is also consistent with the fact that the energy difference between FM and AFM states increases with tensile strain increasing [see Fig. \ref{wh2}(c)]. On the other side, the compressive stain has a negative influence on the exchange coupling between V atoms. The $T_{\rm BKT}$ of monolayer V$_2$PN and V$_2$AsN decreased quickly with the magnitude of compressive strain increasing. Compared to the $T_{\rm BKT}$, the MAEs did not exhibit regular behavior with the temperatures changing. 
The $E_{MSA}$ decreases as the distance between atoms increases, while the $E_{MCA}$ has the opposite tendency [see Fig. S9]. Especially, the orbital resolved MAE shows that the contribution of hybridization $\left\langle {d^{+}_{x^2-y^2}} \right|{\hat L_z}\left| d^{-}_{xy} \right\rangle $ is the main
reason for such a irregular behavior [see Fig. S9]. 
The MAE of monolayer V$_2$PN and V$_2$AsN ranges from $271.3 \sim 321.9$ $\mu$eV per V atom and $88.5 \sim 141.1$ $\mu$eV per V atom under strain ranging from -4\% to 5\%, respectively. In general, with a larger MAE and a higher T$_{BKT}$, monolayer V$_2$AsN is better than V$_2$PN for the material realization of the high-temperature 2D-XY ferromagnetism. 

\subsection{Acknowledgments}
In summary, based on first-principles calculations,
we systemically investigated the structural and magnetic properties of Janus monolayer V$_{2}$XN (X = P and As).
The strong electron correlation effect of $d$ orbital of the V atoms was considered in both structure relaxation and electron structures.
Monolayer V$_{2}$XN has a square lattice with good structural stability and FM half-metallic characters. 
The MAE is 116.6 $\mu$eV/V atom and 292.0 $\mu$eV/V atom for monolayer V$_{2}$PN and V$_{2}$AsN, respectively.
Both the magnetocrystalline anisotropy and shape anisotropy favor an in-plane magnetization, leading to an easy magnetization $xy$-plane in monolayer V$_{2}$XN.  
MC simulations based on the 2D XXZ model indicated that monolayer V$_{2}$XN occurs the BKT phase transition with the emergence of topological non-trivial meron and antimeron spin textures. Through the analysis of the critical behavior of in-plane susceptibility, the $T_{\rm BKT}$ of monolayer V$_{2}$PN and V$_{2}$AsN was estimated as 395 K and 434 K, respectively. A 5\% tensile strain can increase the corresponding $T_{\rm BKT}$ to 508.7 K and 531.3 K, respectively. 
Our work indicated that V-based Janus structures were candidate materials to realize the high-temperature 2D-XY ferromagnetism for spintronic applications.

\clearpage
\bibliographystyle{unsrt}
\bibliography{vpas}

\end{document}